\documentclass[superscriptaddress, aps, pra, twocolumn, nofootinbib]{revtex4-2}

\usepackage{dsfont}
\usepackage{upgreek}
\usepackage{amsfonts}
\usepackage{amsmath}
\usepackage{graphicx}% Include figure files
\usepackage{dcolumn}% Align table columns on decimal point
\usepackage{bm}% bold math
\usepackage[colorlinks, allcolors={blue}]{hyperref}% add hypertext capabilities
\usepackage{braket}
\newcommand{\orcid}[1]{\href{https://orcid.org/#1}{\includegraphics[width=7pt]{orcid.png}}}

\begin{document}

\title{Coherence and realism in the Aharonov-Bohm effect}
\author{Ismael L. Paiva}
\email{ismaellpaiva@gmail.com}
\affiliation{H. H. Wills Physics Laboratory, University of Bristol, Tyndall Avenue, Bristol BS8 1TL, United Kingdom}
\affiliation{Faculty of Engineering and the Institute of Nanotechnology and Advanced Materials, Bar-Ilan University, Ramat Gan 5290002, Israel}

\author{Pedro R. Dieguez}
\affiliation{International Centre for Theory of Quantum Technologies, University of Gda\'nsk, Jana Bazynskiego 8, 80-309 Gda\'nsk, Poland}

\author{Renato M. Angelo}
\affiliation{Department of Physics, Federal University of Paran\'a, P.O. Box 19044, 81531-980 Curitiba, Paran\'a, Brazil}

\author{Eliahu Cohen}
\affiliation{Faculty of Engineering and the Institute of Nanotechnology and Advanced Materials, Bar-Ilan University, Ramat Gan 5290002, Israel}

\begin{abstract}
The Aharonov-Bohm effect is a fundamental topological phenomenon with a wide range of applications. It consists of a charge encircling a region with a magnetic flux in a superposition of wave packets having their relative phase affected by the flux. In this work, we analyze this effect using an entropic measure known as realism, originally introduced as a quantifier of a system's degree of reality and mathematically related to notions of global and local quantum coherence. More precisely, we look for observables that lead to gauge-invariant realism associated with the charge before it completes its loop. We find that the realism of these operators has a sudden change when the line connecting the center of both wave packets crosses the solenoid. Moreover, we consider the case of a quantized magnetic-field source, pointing out similarities and differences between the two cases. Finally, we discuss some consequences of these results.
\end{abstract}

\maketitle

\section{Introduction}

The magnetic Aharonov-Bohm (AB) effect~\cite{ehrenberg1949refractive, Aharonov1959} consists of the influence of a magnetic field encircled by a charge in the charge's dynamics, even if it is completely isolated. The phenomenon is typically discussed in terms of a charge $q$ traveling an interferometer with an ideal infinitely long and thin solenoid with an associated magnetic flux $\Phi_B$ at its center. In this case, the charge's state acquires an extra relative phase $\phi_\text{AB} = q\Phi_B/\hbar$ after encircling the interferometer. This effect was experimentally verified multiple times~\cite{chambers1960shift, mollenstedt1962kontinuierliche, tonomura1982observation, olariu1985quantum}, including the noteworthy experiment by Tonomura \textit{et al.} in which the magnetic field source was shielded by a superconductor~\cite{tonomura1986evidence}.

Initially, the AB effect was presented as a concrete manifestation of electromagnetic potentials in quantum theory, giving them a new status in this domain~\cite{Aharonov1959, aharonov1961further}. However, DeWitt~\cite{dewitt1962quantum} objected to this view, showing that the effect could be seen in terms of a nonlocal interaction between the charge and the magnetic field, with no direct reference to potentials~\cite{aharonov1963further}. With this, he defended that the discussion should shift from the role of potentials to local vs. nonlocal theories. DeWitt's article in fact initiated a debate in the literature~\cite{aharonov1962remarks, belinfante1962consequences, aharonov1963further} that has not been settled yet (see, e.g.,~\cite{healey1997nonlocality, santos1999microscopic, choi2004exact, vaidman2012role, kang2015locality, vaidman2015reply, saldanha2016alternative, pearle2017quantum, pearle2017quantized, marletto2020aharonov, saldanha2021local, saldanha2021shielded}). Aharonov himself has changed his view since his seminal paper with Bohm. Now, he describes the effect as a nonlocal interaction between the field and the charge through modular variables~\cite{aharonov2004effect, aharonov2015comment, aharonov2016nonlocality}, a special type of nonlocality dubbed dynamical nonlocality~\cite{aharonov1969modular, aharonov2005quantum}.

In this work, we analyze the AB effect from the perspective of a quantity known as realism, an entropic measure of definiteness of a property $\hat{O}$ of a system (or a subsystem of a larger system) in the state $\hat{\rho}$. This measure was introduced by Bilobran and Angelo~\cite{bilobran2015measure} and entails a generalization of the so-called Einstein-Podolsky-Rosen element of reality~\cite{einstein1935can}. In fact, its construction is based on the idea that a non-selective projective measurement of an observable does not disturb the measured system if the property associated with it is definite, i.e., it is established prior to the measurement. As will be briefly explained in the next section, although this measure comes attached to an ontological interpretation, it mathematically coincides with notions of coherence and, as a result, is closely related to other important information-theoretic quantities often considered in the literature.

Since our work is independent of the choice of interpretation, we use realism as a measure of coherence (or incoherence, as it will be better discussed later), even though it presumably also quantifies the degree of reality of the system. Consequently, at minimum, our work can be seen as an informational treatment of the AB effect, which is typically not the approach used in studies of this effect. In fact, the only work we know that takes a similar route but in a scenario less generic than the one we present here is Ref.~\cite{edet2022quantum}, where the AB effect with Yukawa interaction in the presence of disclination was investigated. More precisely, here, our aim is to analyze the realism of observables before the charge completes the AB loop. For that, we consider both the standard AB effect and the AB effect with a quantized source. In the standard scenario, we look for observables whose realism is gauge invariant and, at the same time, can detect the presence of the magnetic flux. We find a class of such observables. As it will be seen, a peculiar characteristic of their realism is that it suffers an abrupt change when the line connecting the center of both wave packets of the charge crosses the solenoid. This is an example of application of this measure where a sudden discontinuous change is present. In the case of a quantized source, we discuss the role of reference frames and show how the analysis of the standard scenario is modified in this context. We also discuss how our work introduces a useful framework to the study of dynamical nonlocality.

Besides this introduction, the text is organized as follows. In Sec.~\ref{sec:real} we present the definition of realism, review some of its properties, and describe various other definitions of coherence and other quantum correlations that are equivalent or related to it. Then, we start our informational analysis of interferometers. First, in Sec.~\ref{sec:std-int} we consider a standard interferometer, meaning that no magnetic flux is placed on its interior. Following this, in Sec.~\ref{sec:std-ab} we study the standard AB effect, which involves a classical flux. Continuing, in Sec.~\ref{sec:q-ab} we investigate the AB effect with a quantized flux. Later, in Sec.~\ref{sec:mod-var} we discuss how our analysis fits and give insights into the study of modular variables and dynamical nonlocality. Finally, in Sec.~\ref{sec:discussion} we present our closing remarks. Appendixes are used to expand some calculations.

\section{Realism and its relation to coherence and other informational measures}
\label{sec:real}

Let $\hat{O}=\sum_j o_j \hat{O}_j$ be a discrete-spectrum observable acting on $\mathcal{H_S}$. Here $\hat{O}_j = \ket{o_j} \bra{o_j}$ is the projector in the direction of the eigenvectors $\ket{o_j}$, whose eigenvalue is $o_j\in\mathbb{R}$. Moreover, let the system of interest, represented by the density matrix $\hat{\rho}$, be associated with the Hilbert space $\mathcal{H_S\otimes H_R}$. Then the realism of $\hat{O}$ is said to be defined for the state $\hat{\rho}$ if $\Phi_{\hat{O}}(\hat{\rho}) = \hat{\rho}$, where $\Phi_{\hat{O}}(\hat{\rho}) = \sum_j (\hat{O}_j \otimes \mathds{1}) \hat{\rho}(\hat{O}_j \otimes \mathds{1})$ is a projection-valued measure.

Based on this idea, an entropic measure that quantifies the indefiniteness of a property $\hat{O}$ for a given state $\hat{\rho}$ can be introduced as
\begin{equation}
    \mathfrak{I}_{\hat{O}}(\hat{\rho}) \equiv \min_{\hat{\varrho}} S\left(\hat{\rho}||\Phi_{\hat{O}}(\hat{\varrho})\right) = S\left(\Phi_{\hat{O}}(\hat{\rho})\right) - S(\hat{\rho}),
\end{equation}
where $S(\hat{\rho}||\hat{\sigma}) = \text{tr}\left[\hat{\rho}(\log{\hat{\rho}} - \log{\hat{\sigma}})\right]$ is the relative entropy and $S(\hat{\rho}) = -\text{tr}\left(\hat{\rho} \log{\hat{\rho}}\right)$ is the von Neumann entropy, where the base of the logarithm can be chosen by convenience. This measure is known as irrealism. Observe that it depends on the context given by states and observables. Moreover, it satisfies $0 \leq \mathfrak{I}_{\hat{O}}(\hat{\rho}) \leq \log{d_S}$, where $d_S = \dim\mathcal{H}_S$, and it vanishes if and only if $\hat{\rho} = \Phi_{\hat{O}}(\hat{\rho})$. From the definition of irrealism, it is also possible to introduce a measure of realism:
\begin{equation}
    \mathfrak{R}_{\hat{O}}(\hat{\rho}) \equiv \log{d_S} - \mathfrak{I}_{\hat{O}}(\hat{\rho}),
    \label{eq:def-real}
\end{equation}
which quantifies how definite a property $\hat{O}$ is for a given state $\hat{\rho}$. If the realism of an operator is maximal for a given system's state or, equivalently, the associated irrealism vanishes, it is said that the system has an element of reality for the respective property.

Naturally, one can envision the employment of different types of entropy other than the von Neumann entropy to quantify realism or irrealism. In fact, it has recently been shown that other entropic measures can be found to partially satisfy a set of physically motivated axioms of realism~\cite{orthey2022quantum}. However, it turns out that only the measure induced by the von Neumann entropy used in Eq.~\eqref{eq:def-real} is known to respect all the axioms proposed.

Moreover, a metric different from the relative entropy might be used to quantify how close a state under scrutiny is to a state that has its elements of reality well defined. Nevertheless, the use of relative entropy is convenient because it allows insightful connections between realism or irrealism and other information-theoretic quantities. To start, it can be shown that~\cite{bilobran2015measure}
\begin{equation}
    \mathfrak{I}_{\hat{O}}(\hat{\rho}) = \mathfrak{C}_{\hat{O}}(\hat{\rho}_S) + \mathfrak{D}_{\hat{O}}(\hat{\rho}),
    \label{eq:iid}
\end{equation}
where $\mathfrak{C}_{\hat{O}}(\hat{\rho}_S) \equiv S\left(\Phi_{\hat{O}}(\hat{\rho}_S)\right) - S(\hat{\rho}_S)$ is the coherence of $\hat{\rho}_S$ on the eigenbasis of $\hat{O}$~\cite{baumgratz2014quantifying}, $\mathfrak{D}_{\hat{O}}(\hat{\rho}) \equiv I_{S:R}(\hat{\rho}) - I_{S:R}(\Phi_{\hat{O}}(\hat{\rho}))$ is the non minimized quantum discord, and $I_{S:R}(\hat{\rho}) = S(\hat{\rho}_S) + S(\hat{\rho}_R) - S(\hat{\rho})$ is the quantum mutual information. Clearly, for completely separable states the irrealism reduces to the amount of quantum coherence, showing that coherence is sufficient to preclude classical reality~\cite{orthey2022quantum}.

In addition, from the strong subadditivity of the von Neumann entropy, it holds that $S(\hat{\rho}) + S(\Phi_{\hat{O}}(\Phi_{\hat{O}'}(\rho)))\leq S(\Phi_{\hat{O}}(\hat{\rho})) + S(\Phi_{\hat{O}'}(\hat{\rho}))$  for a pair of observables $\hat{O}$ and $\hat{O}'$ associated with mutually unbiased bases (MUBs). Since, in this case, $S(\Phi_{\hat{O}}(\Phi_{\hat{O}'}(\hat{\rho}))) = S(\mathds{1}/d_S \otimes \hat{\rho}_R) = \log{d_S} + S(\hat{\rho}_R)$, it follows that~\cite{freire2019quantifying, dieguez2022experimental}
\begin{equation}
    \label{eq:complementarity}
    \mathfrak{R}_{\hat{O}}(\hat{\rho}) + \mathfrak{R}_{\hat{O}'}(\hat{\rho}) \leq \log{d_S} + S(\hat{\rho}_S) - I_{S:R}(\hat{\rho}).
\end{equation}
Then, correlations between two subsystems deny quantum systems from reaching full realism for a pair of maximally incompatible observables. Moreover, for pure bipartite states, this upper bound reduces to $\log{d_S} - \mathfrak{E}(\hat{\rho})$, where $\mathfrak{E}(\hat{\rho}) = S(\hat{\rho}_A)$ is the entanglement entropy.

Realism was experimentally assessed with nuclear magnetic resonance techniques in the context of Wheeler's delayed-choice quantum experiments~\cite{dieguez2022experimental} and with photonic weak measurements~\cite{mancino2018information} in the study of the emergence of realism upon monitoring~\cite{dieguez2018information, dieguez2018weak}. This measure was also employed in the theoretical study of several other concepts and frameworks, such as nonlocality~\cite{gomes2018nonanomalous,gomes2019resilience,fucci2019tripartite,gomes2022realism}, quantum walks~\cite{orthey2019nonlocality}, resource theory~\cite{costa2020information}, monitoring under weak measurements~\cite{basso2022reality}, and continuous-variable systems (via operational discretization)~\cite{freire2019quantifying} with an application to matter-wave interferometry~\cite{lustosa2020irrealism}.

To point out more connections between realism and other concepts from quantum information, we first note that, had only information about system $S$ been available, i.e., if the state of the system was $\hat{\rho}_S \equiv \text{tr}_R(\hat{\rho})$, the irrealism $\mathfrak{I}_{\hat{O}}(\hat{\rho}_S)$ of $\hat{O}$ associated with $\hat{\rho}_S$ would reduce to the coherence $\mathfrak{C}_{\hat{O}}(\hat{\rho}_S)$, also known as relative entropy of superposition~\cite{aberg2006quantifying}. This quantity was also shown to correspond to the distillable coherence of $\hat{\rho}_S$~\cite{winter2016operational}. More generally, $\mathfrak{I}_{\hat{O}}(\hat{\rho})$ was dubbed quantum-incoherent relative entropy in Ref.~\cite{chitambar2016assisted}, where its role in assisted coherence distillation was investigated, and measurement-dependent thermal discord in Ref.~\cite{jimenez2019quantum}. Irrealism can also be seen as a type of $G$-asymmetry in the context of frames of reference~\cite{vaccaro2008tradeoff, gour2009measuring}.

It should be noted that works related to coherence tend to consider a single fixed basis. In the literature of realism, however, individual studies often consider multiple bases, since each basis is more directly identified with a given property of the system. This is the main reason for the preference of phrasing our results in terms of realism here.

\section{Standard interferometers}
\label{sec:std-int}

\begin{figure}
    \centering
    \includegraphics[width=\columnwidth]{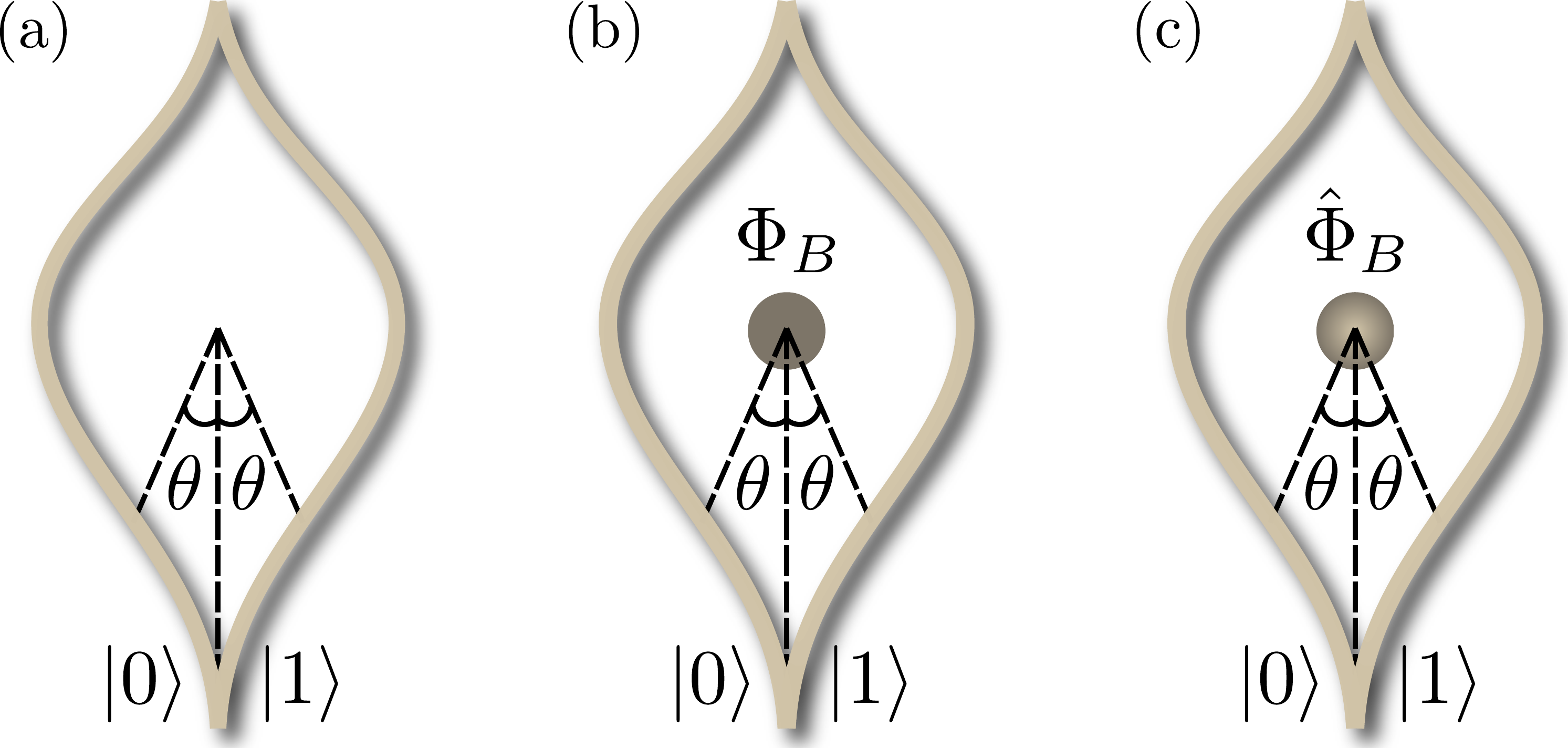}
    \caption{Three interferometers studied with entropic quantities in this work. (a) First, a standard interferometer, where a system travels in a superposition of wave packets in the left and right arms, represented by the states $\ket{0}$ and $\ket{1}$, respectively, is considered. (b) Then the analysis moves to the case where a source of magnetic flux $\Phi_B$, say, a solenoid, is added to the center of the interferometer. (c) Finally, the classical magnetic flux $\Phi_B$ is replaced by a quantized flux $\hat{\Phi}_B$, proportional to the angular momentum $\hat{L}_z$ of a rotating cylinder.}
    \label{fig:interferometers}
\end{figure}

Consider the interferometer represented in Fig.~\ref{fig:interferometers}(a). A quantum particle, associated with the Hilbert space $\mathcal{H}_S$, travels the interferometer in a superposition of wave packets, one in the left arm, represented by the state $|0\rangle$, and the other in the right arm, represented by $|1\rangle$. For now, we do not assume the particle to be a charge. Moreover, for simplicity, we assume the initial state of the particle as soon as it enters the loop is
\begin{equation}
    |\psi_i\rangle = \frac{1}{\sqrt{2}} \left(|0\rangle + |1\rangle\right).
\end{equation}
While in the loop, its state can be represented by
\begin{equation}
    |\psi_0(\theta)\rangle = \frac{1}{\sqrt{2}} \left[|0\rangle + e^{i f(\theta)} |1\rangle\right],
    \label{eq:state-system}
\end{equation}
where $f(\theta)$ is the relative phase between the wave packets after each wave packet traveled an angle $\theta$ in opposite arms. It may include, for instance, effects of phase shifters or other local potentials. In the introduced notation, $\theta$ is $0$ when the wave packets enter the loop and grows monotonically until $\pi$, when the packets interfere.

In the case of a system in a pure state, like the one of interest here, it can be checked that $\mathfrak{I}_{\hat{O}}(\hat{\rho}) = S\left(\Phi_{\hat{O}}(\hat{\rho})\right)$ and, hence, $\mathfrak{R}_{\hat{O}}(\hat{\rho}) = \log d_S - h(\vec{O})$, where $h(\vec{O}) \equiv - \text{tr} \left(\sum_k O_k \log O_k \right)$ is the Shannon entropy with $\vec{O}= (O_k)$ the vector whose components are the eigenvalues of $\Phi_{\hat{O}}(\hat{\rho})$. Since we are analyzing two-level systems, we can work with base-$2$ logarithms and write
\begin{equation}
    \mathfrak{R}_{\hat{O}}(\hat{\rho}) = 1 - h(\lambda),
    \label{eq:def-real2}
\end{equation}
where the notation has been simplified with $h(\lambda) \equiv -\lambda \log_2\lambda -(1-\lambda) \log_2(1-\lambda)$.

With this, observe that the realism associated with $\hat{\sigma}_z$ while the system is inside the interferometer is insensitive to the relative phase. In fact, writing $\hat{\rho}_S \equiv |\psi_0(\theta)\rangle \langle \psi_0(\theta)|$ and observing that $\Phi_{\hat{\sigma}_z}(\hat{\rho}_S) = \mathds{1}/2$, we obtain $\mathfrak{R}_{\hat{\sigma}_z}(\hat{\rho}_S) = 0$. This is not a surprise since, in our analysis, $\hat{\sigma}_z$ is associated with local properties on each arm (its eigenstates correspond to the particle being in a given arm) and the relative phase, as the name already says, is a nonlocal property, in the sense of being a relative quantity between the arms.

However, the realism of, say, $\hat{\sigma}_x$ is sensitive to the relative phase. In fact, as shown in Appendix~\ref{app:std-int}, it is given by Eq.~\eqref{eq:def-real2} with $\lambda = \{1 + \cos[f(\theta)]\}/2$.

More generally, we can define the operator
\begin{equation}
    \hat{\sigma}_g(\theta) \equiv e^{-ig(\theta)} |0\rangle\langle 1| + e^{ig(\theta)} |1\rangle\langle 0|
    \label{eq:gen-xy-op}
\end{equation}
for an arbitrary real function $g$. In this case, Appendix~\ref{app:std-int} shows that $\mathfrak{R}_{\hat{\sigma}_g}(\hat{\rho}_S)$ is given by Eq.~\eqref{eq:def-real2} with $\lambda = \{1 + \cos[f(\theta) - g(\theta)]\}/2$, which, generally, is also a function of the relative phase. Observe that $g\equiv 0$ gives $\hat{\sigma}_x$ and $g\equiv \pi/2$ gives $\hat{\sigma}_y$.

\begin{figure*}
    \centering
    \includegraphics[width=\textwidth]{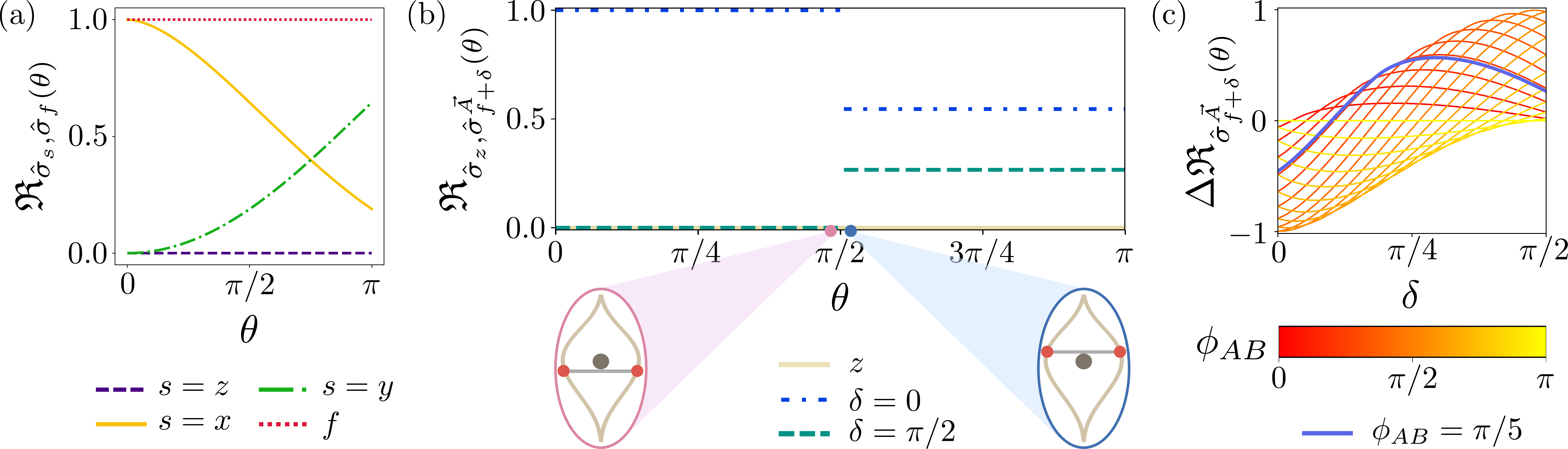}
    \caption{Realism of selected observers in the standard and Aharonov-Bohm interferometers. (a) In a standard interferometer, relative phases can be acquired along the way. The state of the system after its wave packets traveled an angle $\theta$ on each arm is $(\ket{0} + e^{if(\theta)} \ket{1})/\sqrt{2}$. In this case, the realism of $\hat{\sigma}_z$, which is associated with the definiteness of position, vanishes. Also, the realism of $\hat{\sigma}_x$, $\hat{\sigma}_y$, and combinations thereof generally depends on the relative phase $f(\theta)$. However, special combinations, like the operator $\hat{\sigma}_{f+\delta}(\theta)$, where $\delta\in[0,\pi/2]$, constructed from the definition in Eq.~\eqref{eq:gen-xy-op}, are associated with a constant realism throughout the interferometer. The graph shows the behavior of the realism of these observables as a function of $\theta$ for the case where $f(\theta)=\theta/3$ and $\delta=0$. (b) In the AB interferometer with a classical flux, the relative phase gains an extra gauge-dependent term. As a result, the realism of $\hat{\sigma}_x$, $\hat{\sigma}_y$, and their combination is typically a quantity without physical meaning. However, the realism of $\hat{\sigma}_g^{\vec{A}}(\theta)$ defined in Eq.~\eqref{eq:main-operator} is gauge independent. Notably, there is a discontinuous change in the realism of operators of this type once the (virtual) line connecting the center of both wave packets crosses the localized flux. The graph shows the realism of $\hat{\sigma}_z$, $\hat{\sigma}_{x}^{\vec{A}}(\theta) \equiv \hat{\sigma}_{f}^{\vec{A}}(\theta)$, and $\hat{\sigma}_{y}^{\vec{A}}(\theta) \equiv \hat{\sigma}_{f+\pi/2}^{\vec{A}}(\theta)$ in the case where the AB phase is $\phi_\text{AB} = \pi/5$. (c) Analysis of the difference of the realism of each $\hat{\sigma}_{f+\delta}^{\vec{A}}(\theta)$ after the line crosses the flux by its value before that reveals the pattern associated with these jumps.}
    \label{fig:realism-graphs}
\end{figure*}

Also, if $g=f+\delta$, where $\delta$ is a constant, the realism does not depend on the relative phase $f$. It is constant inside the interferometer. The case $\delta=0$ leads to $\mathfrak{R}_{\hat{\sigma}_f(\theta)}(\hat{\rho}_S) = 1$. In Fig.~\ref{fig:realism-graphs}(a), this behavior is juxtaposed with the realism of $\hat{\sigma}_x$, $\hat{\sigma}_y$, and $\hat{\sigma}_z$ for $f(\theta) = \theta/3$. In Ref.~\cite{dieguez2022experimental}, an operator corresponding to $\hat{\sigma}_f(\theta)$ was called a wave operator in contrast to $\hat{\sigma}_z$, which was named a particle operator.

Moreover, it is notable that, in a sense, each operator in the family $\hat{\sigma}_g(\theta)$ indexed by functions $g$ is a counterpart of the modular momentum~\cite{aharonov1969modular, aharonov2005quantum} (see also Sec. II of Ref.~\cite{paiva2022dynamical} for a short introduction to the concept), which is an example of an observable associated with dynamical nonlocality, briefly mentioned in the Introduction. Simply put, this quantity is defined as $\hat{P}_\text{mod} \equiv \hat{P} \mod \hbar/L$ for a certain length $L\in\mathbb{R}^+$. However, periodic functions of it with periods associated with its modularity, which are also modular variables, are the most commonly studied modular operators because they are mathematically simpler since they allow the replacement of $\hat{P}_\text{mod}$ by $\hat{P}$ in their argument. For instance, $e^{i \hat{P}_\text{mod} L/\hbar} = e^{i \hat{P} L/\hbar}$ and its real and imaginary parts are Hermitian modular variables. Then, even in the standard treatment of these variables, a multitude of modular operators can be associated with modular momentum (or functions thereof), similarly to the family $\hat{\sigma}_g(\theta)$.

Modular variables were introduced in a search for operators whose expectation values are a function of relative phases of the wavefunction~\cite{aharonov1969modular, aharonov2005quantum}. Modular momentum, in particular, is suitable to the study of interference phenomena because, if $L$ corresponds to the separation between two wave packets in superposition, its expectation value is a function of the relative phase accumulated by them~\cite{aharonov2017finally}. Because operators in the family $\hat{\sigma}_g(\theta)$ satisfy this central property, we refer to them as counterparts of the modular momentum.

\section{AB effect with a classical magnetic flux}
\label{sec:std-ab}

Consider now the scenario just studied with two key differences: The system traveling in the interferometer is assumed to be a charge and, moreover, a solenoid perpendicular to the interferometer plane is placed at its center, as represented in Fig.~\ref{fig:interferometers}(b). In this case, the AB effect adds a contribution to the relative phase between the superposed wave packets of the charge in such a way that its state after the wave packets are separated by an angle $\theta$ is
\begin{equation}
    |\psi(\theta)\rangle = \frac{1}{\sqrt{2}} \left(|0\rangle + e^{i [f(\theta) + w(\theta)]} |1\rangle\right),
\end{equation}
where
\begin{equation}
    w(\theta) \equiv \frac{q}{\hbar} \left(\int_{\gamma_1(\theta)} \vec{A}\cdot d\vec{s} - \int_{\gamma_0(\theta)} \vec{A} \cdot d\vec{s}\right)
\end{equation}
with $\gamma_0$ and $\gamma_1$ the paths associated with the left and right arms, respectively. The phase $w$, however, depends on the choice of gauge. Then, based on the discussion of the preceding section, it is immediate that the realism of $\hat{\sigma}_z$ still vanishes. Moreover, the realism of $\hat{\sigma}_g$ is given by Eq.~\eqref{eq:def-real2} with $\lambda = \{1 + \cos[(f(\theta) + w(\theta) - g(\theta)]\}/2$ and, therefore, is generally a gauge-dependent quantity. This leads to the conclusion that this observable does not provide a realism that has a physical meaning, which can be understood as the operator itself not having physical meaning.

However, consider the operator
\begin{equation}
    \begin{aligned}
        \hat{\sigma}_g^{\vec{A}}(\theta) \equiv &\; e^{-i\left[g(\theta) - q\int_{\tau(\theta)} \vec{A}\cdot d\vec{s}/\hbar\right]} |0\rangle\langle 1| \\
          &+ e^{i\left[g(\theta) - q\int_{\tau(\theta)} \vec{A}\cdot d\vec{s}/\hbar\right]} |1\rangle\langle 0|,
    \end{aligned}
    \label{eq:main-operator}
\end{equation}
where $\tau(\theta)$ is a line connecting the wave packet on the right-hand side to the wave packet on the left-hand side when they are separated by an angle $\theta$. The dependence of $\hat{\sigma}_g^{\vec{A}}(\theta)$ on the vector potential is analogous to the fact that, in the presence of electromagnetic systems, the kinematic momentum also depends on this quantity.

We can also justify the choice of path $\tau$. As already mentioned in the case of interferometers without the AB effect, the family of operators $\hat{\sigma}_g(\theta)$ indexed by $g$ is analogous to the modular (canonical) momentum in the standard treatment of the problem with continuous variables. In this scenario, the canonical momentum has a clear physical meaning because it coincides with the kinetic momentum. However, in scenarios where the AB effect is manifest, this equivalence no longer holds. Then the modular kinetic momentum, associated with $\vec{P} - q \vec{A}(\vec{r})$, where $\vec{P} = \hat{P}_x \hat{x} + \hat{P}_y \hat{y}$ in our two-dimensional study, becomes the relevant quantity.

An operator typically considered in this case is $e^{i [\vec{P} - q \vec{A}(\vec{r})]\cdot \vec{L}/\hbar}$, where $\vec{L} = \vec{r}_1 - \vec{r}_0$ is the distance between the wave packet located at $\vec{r}_1$ and the wave packet located at $\vec{r}_0$. The real and imaginary parts of this modular variable are Hermitian operators, but the exponential form is used to simplify calculations. Then, noting that
\begin{equation}
    e^{iq\int_{\gamma(\vec{r})} \vec{A}\cdot d\vec{s}/\hbar} \left(\vec{P}\cdot\vec{L}\right) e^{-iq\int_{\gamma(\vec{r})} \vec{A}\cdot d\vec{s}/\hbar} = \left[\vec{P}-q\vec{A}(\vec{r})\right] \cdot \vec{L}/\hbar
\end{equation}
for an arbitrary curve $\gamma$, it follows that~\cite{aharonov2004effect}
\begin{equation}
    e^{i\left[\vec{P}-q\vec{A}(\vec{r})\right] \cdot \vec{L}/\hbar} = e^{-iq\int_{\uptau(\vec{r})} \vec{A}\cdot d\vec{s}/\hbar} e^{i\vec{P}\cdot\vec{L}/\hbar},
    \label{eq:identity}
\end{equation}
where $\uptau(\vec{r})$ is a line connecting $\vec{r}$ to $\vec{r}+\vec{L}$. In other words, in a generic scenario with continuous variables, the modular kinetic momentum of interest becomes the product of the standard modular canonical momentum by $e^{-iq\int_{\uptau(\vec{r})} \vec{A}\cdot d\vec{s}/\hbar}$. Importantly, the line $\uptau$ originates from the identity in Eq.~\eqref{eq:identity}.

Now that our choice of path $\tau$ has been explained, we can proceed to the analysis of the realism of $\hat{\sigma}_g^{\vec{A}}(\theta)$ for the system of interest. For this, we can again use the results of Appendix~\ref{app:std-int} to conclude that $\mathfrak{R}_{\hat{\sigma}_g^{\vec{A}}(\theta)}(\hat{\rho}_S)$ reduces to the expression in Eq.~\eqref{eq:def-real2} with $\lambda = \{1 + \cos[f(\theta) - g(\theta) + q\Phi_\text{enc}/\hbar]\}/2$, where
\begin{equation}
    \Phi_\text{enc} \equiv \int_{\gamma_1(\theta)} \vec{A}\cdot d\vec{s}-\int_{\gamma_0(\theta)} \vec{A}\cdot d\vec{s} + \int_{\tau(\theta)} \vec{A}\cdot d\vec{s}
\end{equation}
refers to the flux encircled by the closed path $\tau + \gamma_1 - \gamma_0$.

Observe that, before $\tau$ crosses the solenoid, i.e., for $\theta < \pi/2$, $\Phi_\text{enc}=0$. Hence, $\lambda = \{1 + \cos[f(\theta) - g(\theta)]\}/2$. However, for $\theta > \pi/2$, $\lambda = \{1 + \cos[f(\theta) - g(\theta) + \phi_\text{AB}]\}/2$. This means that in general there exists a discontinuous change in $\mathfrak{R}_{\hat{\sigma}_g^{\vec{A}}(\theta)}(\hat{\rho}_S)$ once the line connecting the center of the wave packets crosses the solenoid, as illustrated in Fig.~\ref{fig:realism-graphs}(b).

Like in the study of the standard interferometer, the subset of $\{\hat{\sigma}_g^{\vec{A}}(\theta)\}$ characterized by $g=f+\delta$ is of special interest. While the realism of these operators still depends on the relative phase of the charge's state, they depend only on the portion added by the AB effect. Then they do not change inside the interferometer, except when the line connecting the wave packets crosses the flux. In particular, in the case $\delta=0$, for which we define $\hat{\sigma}_x^{\vec{A}}(\theta) \equiv \hat{\sigma}_f^{\vec{A}}(\theta)$, it follows that $\mathfrak{R}_{\hat{\sigma}_x^{\vec{A}}(\theta)}(\hat{\rho}_S) = 1$ for $\theta<\pi/2$ and the expression in Eq.~\eqref{eq:def-real2} with $\lambda = [1 + \cos(\phi_\text{AB})]/2$ for $\theta>\pi/2$. In other words, the realism of $\hat{\sigma}_x^{\vec{A}}(\theta)$ starts with maximal value before the line $\tau$ crosses the solenoid and, generally, drops after that. Moreover, if $\delta=\pi/2,$ the realism operator $\hat{\sigma}^{\vec{A}}_y(\theta) \equiv \hat{\sigma}_{f+\pi/2}^{\vec{A}}(\theta)$ has an opposite (but not complementary) behavior of the realism of $\hat{\sigma}_x^{\vec{A}}(\theta)$. In fact, it vanishes for $\theta < \pi/2$ and has a sudden increase to the value in Eq.~\eqref{eq:def-real2} with $\lambda = [1 - \sin(\phi_\text{AB})]/2$ after that. These behaviors are represented in Fig.~\ref{fig:realism-graphs}(b) for $\phi_\text{AB} = \pi/5$.

We can also compute the change associated with the realism jump of each operator $\{\hat{\sigma}_{f+\delta}^{\vec{A}}(\theta)\}$:
\begin{equation}
    \Delta\mathfrak{R}_{\hat{\sigma}_{f+\delta}^{\vec{A}}(\theta)}(\hat{\rho}_S) = h\left(\frac{1 + \cos(\delta)}{2}\right) - h\left(\frac{1 + \cos(\phi_\text{AB}-\delta)}{2}\right).
\end{equation}
This quantity, graphed in Fig.~\ref{fig:realism-graphs}(c), consists of the difference between the realism after $\theta=\pi/2$ and before that. Observe that it vanishes for $\delta = \phi_{AB}/2$.

\section{AB effect with a quantum magnetic flux}
\label{sec:q-ab}

Now, we consider the scenario depicted in Fig.~\ref{fig:interferometers}(c), which corresponds to the AB effect with a quantized source of magnetic field. This type of study, where the source or the field gets a quantum treatment, is of vast interest in the literature~\cite{peshkin1961quantum, aharonov1991there, santos1999microscopic, choi2004exact, vaidman2012role, pearle2017quantum, pearle2017quantized, li2018transition, marletto2020aharonov, horvat2020probing, saldanha2021shielded}. We follow the approach introduced in Ref.~\cite{aharonov1991there} and studied in post-selected scenarios in Ref.~\cite{paiva2021aharonov}, where the source of the magnetic field is an infinitely long cylindrical shell with a moment of inertia $I_c$ and angular momentum $\hat{L}_z$ (associated with the Hilbert space $\mathcal{H}_R$). Thus, in this configuration, the cylinder is analogous to an infinite solenoid.

The Hamiltonian of such a cylinder and a charge $q$ with mass $m$ and moment of inertia $I_q$ moving in the $xy$ plane can be furnished as~\cite{aharonov1991there}
\begin{equation}
    \hat{H} = \frac{1}{2m} \hat{P}_r^2 + \frac{1}{2I_q} \left(\hat{P}_\theta - \frac{qK}{2\pi} \hat{L}_z\right)^2 + \frac{1}{2I_c} \hat{L}_z^2,
    \label{eq-hamilt}
\end{equation}
where $K$ is a constant inversely proportional to $I_c$ and $P_r$ and $P_\theta$ are, respectively, the canonical radial and angular momentum of the charge. Observe that the standard vector potential is replaced by an interaction term that contains the operator vector potential $\vec{A} = (K/2\pi r) \hat{L}_z \hat{\theta}$, which is similar in form to the Coulomb or Lorenz gauge. Also, the associated magnetic flux is $\hat{\Phi}_B = K \hat{L}_z$. In these expressions, $r$ is the distance between a point and the center of the cylinder and $\hat{\theta}$ is the unit vector in the angular direction, and not an operator.

One may ask how this Hamiltonian transforms under the choice of a different gauge. This was the main question investigated in Ref.~\cite{aharonov1991there}. For completeness, we briefly summarize the authors' approach and results. Starting with a classical Lagrangian treatment, they observed that non-trivial changes of gauges add a term to the Lagrangian that depends on the second time derivative of the angular variable $\eta$ of the cylinder. Because of this and in order to obtain the Hamiltonian of the joint system, a new variable corresponding to the first time derivative of $\eta$ is introduced. This new variable and $\eta$ are treated as independent in the Legendre transform. The resultant Hamiltonian then should satisfy a constraint. This constraint can be solved either before the quantization of the system (with a process known as reduced quantization) or after that (with a method known as Dirac quantization)~\cite{dirac1964lectures, teitelboim1992quantization, vanrietvelde2020change}. The authors used reduced quantization and observed that, in the final version of the Hamiltonian, the charge and the cylinder are not separable systems. More specifically, the momentum of the cylinder turned out to be the momentum conjugated to a coordinate resultant from a linear combination of $\eta$ and $\theta$. Then, different gauges should be associated with different choices of classical coordinates to describe the system. With this in mind, we assume the usual coordinates in our treatment, i.e., $\eta$ and $\theta$, which lead to a mathematically simpler Hamiltonian with the operator vector potential written in the Coulomb gauge.

A generic angular momentum state of the cylinder can be written as $|\xi\rangle = \sum_{m_\ell\in\Gamma_\ell} c_{m_\ell} |m_\ell\rangle$, where $\Gamma_\ell \equiv \{-\ell, -\ell+1, \hdots, 0, \hdots, \ell-1, \ell\}$, $c_{m_\ell}$ are complex coefficients, and $|m_\ell\rangle$ is an eigenstate of $L_z$ with eigenvalue $m_\ell \hbar$. Then the state of the joint system while the charge is inside the interferometer is
\begin{equation}
    \ket{\Psi(\theta)} = \sum_{m_\ell\in\Gamma_\ell} \frac{c_{m_\ell} e^{ig_{m_\ell}(\theta)}}{\sqrt{2}} (\ket{0} + e^{i[f(\theta) + Kqm_\ell \theta/\pi]} \ket{1}) \otimes \ket{m_\ell},
    \label{map}
\end{equation}
where $g_{m_\ell}$ is related to a global phase due to the AB effect associated with $m_\ell$ and, moreover, the free evolution of the cylinder. We also define $\hat{\rho} \equiv \ket{\Psi (\theta)} \bra{\Psi (\theta)}$. Then the reduced state of the charge can be written as
\begin{equation}
    \begin{aligned}
        \hat{\rho}_S = \frac{1}{2} \left[ \mathds{1}_S + \sum_{m_\ell \in \Gamma_\ell} |c_{m_\ell}|^2 \right. & \left(e^{-i [f(\theta) + Kq m_\ell \theta/\pi]} \ket{0} \bra{1} \right. \\
            &\left.\left. + e^{i [f(\theta) + Kq m_\ell \theta/\pi]} \ket{1} \bra{0} \right) \right],
    \end{aligned}
\end{equation}
which corresponds to a proper statistical mixture of a system affected by vector potentials $\vec{A}_{m_\ell} = (K \hbar m_\ell/2\pi r) \hat{\theta}$, which have an associated magnetic flux $\Phi_B^{m_\ell} = K \hbar m_\ell$.

\begin{figure}
    \centering
    \includegraphics[width=\columnwidth]{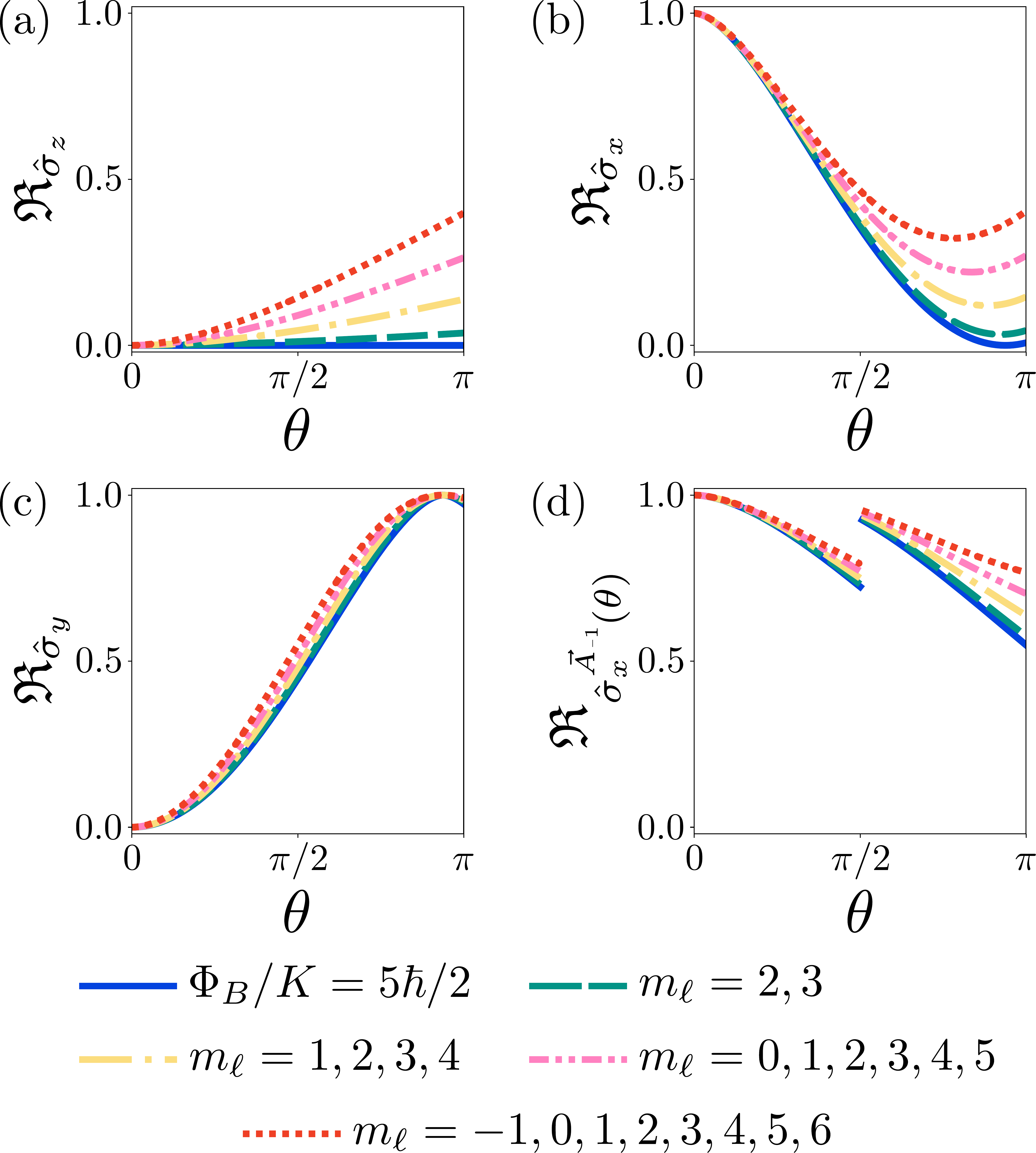}
    \caption{Lack of coherence of the charge's reduced state in the eigenbases of selected observables. Since this quantity corresponds to the realism of these observables had only information about the charge been known, we refer to it as the realism of the charge. The solid line represents the scenario with a classical flux $\Phi_B = 5K\hbar/2$ and $\phi_\text{AB} = \pi/5$. The other curves correspond to scenarios with a quantized flux, associated with the angular momentum $\hat{L}_z$ of a cylinder with $\ell \geq 6$. While the average initial flux in all cases is also $\Phi_B$, each preparation is given by distinct even superpositions of the indicated subset of eigenstates of $\hat{L}_z$. For instance, $m_\ell = 2,3$ corresponds to the cylinder being prepared in the state $(\ket{2}+\ket{3})/\sqrt{2}$. (a) Entanglement between the particle and the cylinder increases the realism of $\hat{\sigma}_z$, i.e., the realism of $\hat{\sigma}_z$ after the cylinder is traced out. The same analysis holds for the realism of (b) $\hat{\sigma}_x$ and (c) $\hat{\sigma}_y$. (d) Furthermore, in the case with a quantized flux, it is not possible to find local operators that are analogous to the $\hat{\sigma}_{f+\delta}^{\vec{A}}(\theta)$ previously considered. Instead, one can only construct a similar operator associated with each $m_\ell$ or linear combinations of them. The graph shows the realism of the operator associated with $m_\ell=-1$ as a function of $\theta$.}
    \label{fig:ab-quantum-local}
\end{figure}

Since $\Phi_{\hat{\sigma}_z}(\hat{\rho}_S) = \mathds{1}_S/2$, the realism\footnote{Strictly speaking, this quantity is the incoherence of $\hat{\rho}_S$ in a given basis. However, since it equates to the realism of the associated operator had only system $S$ been known, we simply refer to it as realism of the associated operator for $\hat{\rho}_S$.} associated with $\hat{\sigma}_z$, when ignoring the cylinder, is $\mathfrak{R}_{\hat{\sigma}_z}(\hat{\rho}_S) = h(\lambda_0)$, with
\begin{equation}
    \lambda_0 \equiv \frac{1}{2} \left[1 - \left| \sum_{m_\ell \in \Gamma_\ell} |c_{m_\ell}|^2 e^{-i [f(\theta) + Kq m_\ell \theta/\pi]} \right| \right].
\end{equation}
Observing that the entanglement entropy of $\hat{\rho}$ is $\mathfrak{E}(\hat{\rho}) = h(\lambda_0)$, we note that the realism of $\hat{\sigma}_z$ for $\hat{\rho}_S$ reduces to this quantity. In other words, due to correlations between the charge and the cylinder, there is an increase in the realism of $\hat{\sigma}_z$ associated with $\hat{\rho}_S$ relative to the interferometer the AB effect with a classical flux. This is represented in Fig.~\ref{fig:ab-quantum-local}(a).

Observe that this realism has a dependence on a phase that is related to the Coulomb gauge. In fact, as we have mentioned, the choice of coordinates fixed the electromagnetic gauge. This result is consistent with the fact that entanglement is reference-frame dependent~\cite{aharonov1984quantum, angelo2011physics, giacomini2019quantum}.

Similarly, we conclude that the realism associated with $\hat{\sigma}_x$ is $\mathfrak{R}_{\hat{\sigma}_x}(\hat{\rho}_S) = 1 + h(\lambda_0) - h(\lambda)$, with
\begin{equation}
    \lambda = \frac{1}{2} \left[1 - \sum_{m_\ell\in\Gamma_\ell} |c_{m_\ell}|^2 \cos(f(\theta) + Kqm_\ell \theta/\pi)\right].
\end{equation}
Neglecting the term $h(\lambda_0)$, this is closely related to the result we obtained with a classical source. However, here, we repeat, the choice of coordinates fixed the electromagnetic gauge. This is illustrated in Fig.~\ref{fig:ab-quantum-local}(b). See also Fig.~\ref{fig:ab-quantum-local}(c) for the realism of $\hat{\sigma}_y$.

Moreover, it can be verified that
\begin{equation}
    \int_{\tau(\theta)} \vec{A}_{m_\ell} \cdot d\vec{s} = K \hbar m_\ell \pi^{-1} (\pi \Theta(\theta-\pi/2) - \theta),
\end{equation}
where $\Theta$ denotes the Heaviside step function. As a result, for each $m_\ell$, the counterpart of the operator $\hat{\sigma}_x^{\vec{A}}(\theta)$ is
\begin{equation}
    \begin{aligned}
        \hat{\sigma}_x^{\vec{A}_{m_\ell}}(\theta) = & \; e^{-i\left\{f(\theta) - Kq m_\ell [\pi \Theta(\theta-\pi/2) - \theta]/\pi\right\}} \ket{0} \bra{1} \\
           & + e^{i\left\{f(\theta) - Kq m_\ell [\pi \Theta(\theta-\pi/2) - \theta]/\pi\right\}} \ket{1} \bra{0}.
    \end{aligned}
\end{equation}
An example of realism of this class of operators can be seen in Fig.~\ref{fig:ab-quantum-local}(d).

However, when restricted to the state of system $S$ alone, it is not possible to prepare an operator whose realism does not depend on the choice of coordinates. In fact, this is only possible in the case of weak interactions. This scenario refers to cases where the charge and the cylinder remain at least approximately separable throughout their interaction and, hence,
\begin{equation}
    \begin{aligned}
        \sum_{m_\ell\in\Gamma_\ell} |c_{m_\ell}|^2 &e^{\pm i Kq m_\ell [\pi \Theta(\theta-\pi/2) - \theta]/\pi} \\
            & \hspace{10mm} \approx e^{\pm i Kq \langle \hat{L}_z \rangle [\pi \Theta(\theta-\pi/2) - \theta]/\pi \hbar},
    \end{aligned}
\end{equation}
where $\langle \hat{L}_z \rangle = \hbar \sum_{m_\ell\in\Gamma_\ell} |c_{m_\ell}|^2  m_\ell$. This allows us to define an analog of $\hat{\sigma}_x^{\vec{A}}(\theta)$ independent of $m_\ell$. It includes, in particular, the case in which the cylinder is prepared in an eigenstate of $\hat{L}_z$. A more detailed analysis of the weakly interacting scenario can be found in Ref.~\cite{paiva2021aharonov}. For the present work, it is important to note that $h(\lambda_0)=0$ in this case, which makes every realism identical to their expression in the study of the AB effect with a classical source with magnetic flux $\Phi_B = K \langle \hat{L}_z \rangle$.

\begin{figure}
    \centering
    \includegraphics[width=\columnwidth]{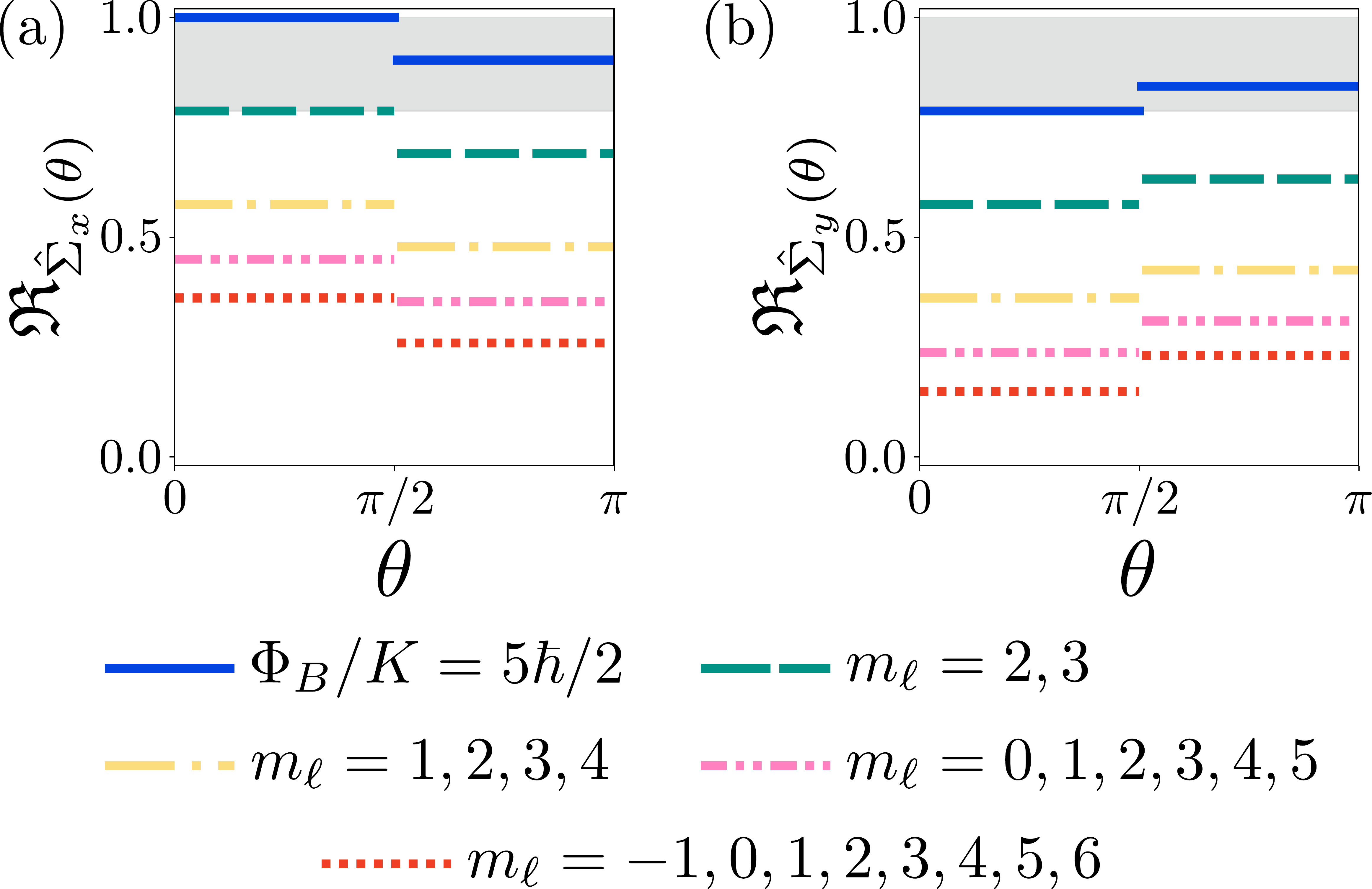}
    \caption{Realism of $\hat{\Sigma}_x(\theta)$ and $\hat{\Sigma}_y(\theta)$ for the joint state of the charge and cylinder in the case of $\ell=6$ and $qK=2\pi/25$. The shaded area represents the region where the realism of observables of the charge alone may lie if considered on the same scale as the realism of observables that are non-trivial in both systems. Their minimum value is $\log_{26}13$ in the illustrated example. The operators $\hat{\Sigma}_x(\theta)$ and $\hat{\Sigma}_y(\theta)$ are analogous to $\hat{\sigma}_{x}^{\vec{A}}(\theta) \equiv \hat{\sigma}_{f}^{\vec{A}}(\theta)$ and $\hat{\sigma}_{y}^{\vec{A}}(\theta) \equiv \hat{\sigma}_{f+\pi/2}^{\vec{A}}(\theta)$, respectively, in the case of a classical flux.}
    \label{fig:ab-quantum-global}
\end{figure}

In the general case, however, we can define a counterpart of $\hat{\sigma}_x^{\vec{A}}(\theta)$ as an operator that acts on both systems $S$ and $R$. More precisely,
\begin{equation}
    \hat{\Sigma}_x(\theta) \equiv \sum_{m_\ell\in\Gamma_\ell} \hat{\sigma}_x^{\vec{A}_{m_\ell}}(\theta) \otimes \ket{m_\ell} \bra{m_\ell}.
    \label{eq:def-sigma-quant}
\end{equation}
The realism of this and other similar operators can be computed using base-$(4\ell + 2)$ logarithms as $\mathfrak{R}_{\hat{O}}(\hat{\rho}) = 1 - \mathfrak{I}_{\hat{O}}(\hat{\rho})$. Then, following the derivation in Appendix~\ref{app:q-ab-x}, we conclude that, for $\theta < \pi/2$, $\mathfrak{R}_{\hat{\Sigma}_x(\theta)}(\hat{\rho}) = 1 - h(\vec{C})$, where $\vec{C} = (|c_{m_\ell}|^2)$. Moreover, for $\theta > \pi/2$, $\mathfrak{R}_{\hat{\Sigma}_x(\theta)}(\hat{\rho}) = 1 - h(\vec{C}) - \sum_{m_\ell\in\Gamma_\ell} |c_{m_\ell}|^2 h(\lambda_{m_\ell})$, where $\lambda_{m_\ell} = [1 + \cos(q\Phi_B^{m_\ell}/\hbar)]/2$. This is illustrated in Fig.~\ref{fig:ab-quantum-global}(a).

Similarly, we can define $\hat{\Sigma}_y(\theta)$ as
\begin{equation}
    \hat{\Sigma}_y(\theta) \equiv \sum_{m_\ell\in\Gamma_\ell} \hat{\sigma}_y^{\vec{A}_{m_\ell}}(\theta) \otimes \ket{m_\ell} \bra{m_\ell}.
    \label{eq:sigmay-quant}
\end{equation}
Then a simple adaptation of the computation in Appendix~\ref{app:q-ab-x} leads to $\mathfrak{R}_{\hat{\Sigma}_y(\theta)}(\hat{\rho}) = 1 - \log_{4\ell+2}2 - h(\vec{C})$ if $\theta < \pi/2$ and $\mathfrak{R}_{\hat{\Sigma}_x(\theta)}(\hat{\rho}) = 1 - h(\vec{C}) - \sum_{m_\ell\in\Gamma_\ell} |c_{m_\ell}|^2 h(\lambda_{m_\ell})$ with $\lambda_{m_\ell} = [1 - \sin(q\Phi_B^{m_\ell}/\hbar)]/2$ if $\theta > \pi/2$. See Fig.~\ref{fig:ab-quantum-global}(b) for an explicit example.

Finally, since $\hat{\sigma}_z$ is insensitive to the presence of the magnetic flux, its analog for the joint system is simply $\hat{\sigma}_z \otimes \mathds{1}_R$. Then, for completion, we can also compute the realism of $\hat{\sigma}_z$ associated with the whole $\hat{\rho}$. As shown in Appendix~\ref{app:sigma-z}, when calculated in the same scale used to calculate the realism of $\hat{\Sigma}_x(\theta)$ and $\hat{\Sigma}_y(\theta)$, it can be concluded that $\mathfrak{R}_{\hat{\sigma}_z}(\hat{\rho}) = \log_{4\ell+2}(2\ell+1)$ for every $\theta$, which is the minimum value for a realism of a local observable in this scale, as discussed in Fig.~\ref{fig:ab-quantum-global}. Moreover, when computed in the same scale used in Sec.~\ref{sec:std-ab}, i.e., with base-$2$ logarithms, we obtain $\mathfrak{R}_{\hat{\sigma}_z}(\hat{\rho}) = 0$ for every $\theta$, just like in the AB effect with a classical flux.

\section{Realism and modular variables}
\label{sec:mod-var}

As we have already discussed, the operators $\hat{\sigma}_g(\theta)$ introduced in Sec.~\ref{sec:std-int} are counterparts of the modular variable of (canonical) momentum. This coincides with the modular kinetic momentum when magnetic fields are not taken into consideration. However, when a classical source of magnetic field is at play, the analogs of the modular kinetic momentum become the operators $\hat{\sigma}_g^{\vec{A}}(\theta)$ presented in Sec.~\ref{sec:std-ab}.

The report of the sudden change in the realism of the latter observed here is unusual for this type of behavior in the literature of realism. While, on the one hand, this is a surprising result, it is, on the other hand, a generic feature of topological effects. In fact, discontinuous changes of some relevant quantities are not uncommon to topological effects. Moreover, this characteristic is in line with a previous study of the expectation value of the modular kinetic momentum presented in Ref.~\cite{aharonov2004effect}.

It should be noted that, although modular variables find applications in continuous-variable quantum information processing~\cite{gottesman2001encoding, vernaz2014continuous, ketterer2016quantum}, they were first introduced to explain interference phenomena in the original description of Heisenberg for quantum mechanics. In this description, a quantum particle is assumed to always be localized, even if its location cannot be known. This makes quantum interference a particularly challenging phenomenon to explain. With modular variables, however, an explanation is possible in terms of operators whose Heisenberg dynamics presents some sort of nonlocal dependence on remote points in space~\cite{aharonov1969modular, aharonov2017finally}. Even though this interpretation is not necessary to work with these quantities, we show that it can be conciliated with the analysis of realism.

 To start, we first discuss a form of self-consistency in the study of modular variables. In fact, these variables satisfy a principle known as the complete uncertainty principle~\cite{aharonov2005quantum, aharonov2017finally}, which states that a (continuous and dimensionless) modular variable $\hat{\Lambda}$ is maximally uncertain if and only if the expectation value of $e^{i n \hat{\Lambda}}$ vanishes for every $n\in\mathbb{N}$. This implies that, when information about, say, the position of a particle is obtained, its modular momentum becomes completely uncertain, which can be seen as the reason for the destruction of the interference pattern later on.

This result has an analog for the operators considered here: The expectation value of every operator in the family $\hat{\sigma}_g^{\vec{A}}(\theta)$ [or $\hat{\sigma}_g(\theta)$] vanishes if and only if they are maximally uncertain. In fact, the uncertainty of an operator $\hat{O}$ is $\Delta\hat{O} \equiv \sqrt{\langle\hat{O}^2\rangle - \langle\hat{O}\rangle^2}$ and in the case of the operators we are considering $\hat{O}^2 = \mathds{1}$. As a result, $\Delta\hat{O} = \sqrt{1 - \langle\hat{O}\rangle^2}$, which reaches its maximal value if and only if $\langle\hat{O}\rangle = 0$. Observe that the latter is the case for the operators of the families $\hat{\sigma}_g^{\vec{A}}(\theta)$ or $\hat{\sigma}_g(\theta)$ if and only if $\Phi_{\hat{\sigma}_z}(\hat{\rho}) = \hat{\rho}$.

Now, realism measures the degree of definiteness of a property of the system. Then, in a way, it also quantifies the amount of disturbance that is induced by a measurement of the observable under scrutiny: The greater the realism, the smaller the disturbance~\cite{maccone2007entropic, busch2009no, sharma2018trade, dieguez2018information}. As a result, the complete uncertainty principle for the scenario of interest in this work can be understood as an association between the degree of knowledge about the position of a particle inside an interferometer (i.e., the realism of $\hat{\sigma}_z$) and the vanishing of the realism of the corresponding modular variable [i.e., $\hat{\sigma}_g(\theta)$ or $\hat{\sigma}_g^{\vec{A}}(\theta)$]. This in turn can be seen as a consequence of the complementarity between the realism of $\hat{\sigma}_z$ and $\hat{\sigma}_g(\theta)$ [or $\hat{\sigma}_g^{\vec{A}}(\theta)$]. In fact, using base-$2$ logarithms, Eq.~\eqref{eq:complementarity} leads to
\begin{equation}
    \mathfrak{R}_{\hat{\sigma}_z}(\ket{\psi_S} \bra{\psi_S}) + \mathfrak{R}_{\hat{\sigma}_g^{\vec{A}}(\theta)}(\ket{\psi_S} \bra{\psi_S}) \leq 1
    \label{eq:comp-sigmas}
\end{equation}
for an arbitrary pure state $\ket{\psi_S}$ of a system traveling the interferometer. This means that the realism of both $\hat{\sigma}_z$ and $\hat{\sigma}_g^{\vec{A}}(\theta)$ cannot be simultaneously maximal. Observe that in the particular state of the system we considered, this relation is trivial since $\mathfrak{R}_{\hat{\sigma}_z}$ vanishes. A similar relation holds for the realism of $\hat{\sigma}_z$ and $\hat{\sigma}_g(\theta)$.

Nevertheless, we note that the usual perspective that is taken in the literature of realism is that, whenever irrealism vanishes, there is an element of reality associated with the observable for the system of interest. If this perspective is assumed, the results presented here can be read as a statement that the AB effect implies a discontinuous change in physical reality. However, it should be pointed out that this view implies a philosophical departure from the original picture of Heisenberg for quantum theory, which was the original motivation for the introduction of modular variables. Moreover, it would be interesting to understand how this notion relates to other results on the reality of quantum states. For instance, does the Pusey-Barrett-Rudolph theorem~\cite{pusey2012reality}, which states that quantum states cannot be purely epistemic, have any implications to the usual interpretation of realism? This type of question deserves a deeper analysis in a future investigation, with possible interpretational implications for this work and the entire literature of realism. However, we emphasize that the main results presented here are independent of choices of interpretation.

An aspect that has not been previously highlighted in the literature but is noteworthy is that the complementarity principle, a cornerstone in the study of realism, is one of the consequences of entropic uncertainty relations~\cite{coles2017entropic}. In particular, Eq.~\eqref{eq:comp-sigmas} follows from the fact that $S(\Phi_{\hat{\sigma}_z}(\hat{\rho}_S)) + S(\Phi_{\hat{\sigma}_g^{\vec{A}}(\theta)}(\hat{\rho}_S)) \geq 1$.

More generally, the entropic uncertainty relation associated with an arbitrary pair of observables $\hat{O}$ and $\hat{O}'$ can be written as $S(\Phi_{\hat{O}}(\hat{\rho}_S)) + S(\Phi_{\hat{O}'}(\hat{\rho}_S)) \geq \log{(1/c)}$, where $c\equiv \max_{j,k} |\langle o_j | o_k' \rangle|^2$ with $\ket{o_j}$ and $\ket{o_k'}$ denoting eigenstates of $\hat{O}$ and $\hat{O}'$, respectively~\cite{maassen1988generalized, sanchez1993entropic, wu2009entropic}. In the case of bipartite systems, this relation becomes~\cite{berta2010uncertainty}
\begin{equation}
    S(\Phi_{\hat{O}}(\hat{\rho})) + S(\Phi_{\hat{O}'}(\hat{\rho})) \geq \log{(1/c)} + S_{S:R}(\hat{\rho}),
\end{equation}
where $S_{S:R}(\hat{\rho}) \equiv S(\hat{\rho}) - S(\hat{\rho}_R)$ is the conditional entropy. This implies that
\begin{equation}
    \mathfrak{R}_{\hat{O}}(\hat{\rho}) + \mathfrak{R}_{\hat{O}'}(\hat{\rho}) \leq \log{(d_S^2 c)} + S(\hat{\rho}) + S(\hat{\rho}_R),
    \label{eq:gen-bound}
\end{equation}
which can be seen as a generalization of Eq.~\eqref{eq:complementarity} for an arbitrary pair of observables (and not only pairs associated with MUBs). However, for MUBs, this bound is not as tight as the one in Eq.~\eqref{eq:complementarity} since, in this case, Eq.~\eqref{eq:gen-bound} gives $\mathfrak{R}_{\hat{O}}(\hat{\rho}) + \mathfrak{R}_{\hat{O}'}(\hat{\rho}) \leq \log{d_S} + S(\hat{\rho}) + S(\hat{\rho}_R)$.

Before concluding this section, it should be mentioned that one of the objectives of approaches like the one presented here is that they may lead to a connection between the dynamical nonlocality associated with modular variables and the more common kinematic nonlocality such as the one present in Bell scenarios~\cite{brunner2014bell}. The latter has the benefit of admitting a device-independent formulation leading to bounds that allow their certification. The former, which explicitly depends on the specific dynamical equations of motion, generates more controversy. In our study, however, we have presented an analysis involving modular operators and their relation with other informational quantities. This may help to elucidate their meaning. Moreover, in the case of a quantized flux, the best analogs of the modular kinetic momentum consist of operators of the type defined in Eqs.~\eqref{eq:def-sigma-quant} and \eqref{eq:sigmay-quant}, which is a joint property of the charge and the flux. Using an analogy with controlled gates in quantum circuits, these operators are like properties of the charge controlled by the flux source's state. Then, it seems that there is an interplay between dynamical nonlocality and the more common kinematic nonlocality. Since the AB effect is what generates the entanglement between the systems, one could even state that, apparently, dynamical nonlocality generates kinematic nonlocality in this case. This suggests the possibility of a study on how the former can be used as a resource to generate the latter (and, potentially, vice versa~\cite{elitzur2015quantum}). This investigation is left for future work.

\section{Closing remarks}
\label{sec:discussion}

We have analyzed the AB effect before the charge completes a full loop from a quantum informational perspective, mainly focusing on the realism measure. In the case of a classical flux, we have shown that the operators $\hat{\sigma}_z$ and the family $\hat{\sigma}_g^{\vec{A}}(\theta)$ indexed by functions $g$ are associated with gauge-independent realism. The latter is particularly relevant because these observables are sensitive to the presence of magnetic flux. We have also considered the effect with a quantized flux. In this case, an analog of $\hat{\sigma}_g^{\vec{A}}(\theta)$ can only be found in the case of weak interactions or if the operator includes degrees of freedom of the source. More specifically, in the case of arbitrary interaction strength, the counterpart of $\hat{\sigma}_g^{\vec{A}}(\theta)$ consists of operations controlled by the eigenstates of the angular momentum of the cylinder, such as the operators in Eqs.~\eqref{eq:def-sigma-quant} and \eqref{eq:sigmay-quant}.

Still in the case of a quantized flux, we have observed that the entanglement increases the realism of the local observers that continue to be well-defined in this scenario. This is the case because the entanglement between the two systems is associated with one having some information about the other. In particular, measurements of the flux can partially define the realism of the position of the charge. This is similar to the complementarity between information and realism introduced in Ref.~\cite{dieguez2018information}.

This work promotes the study of modular variables and the concept of dynamical nonlocality from the perspective of quantities more familiar to quantum information theory. This was done with an entropic analysis of counterparts of these variables. Although realism was the chosen quantity, various other coherence measures would lead to equivalent or similar conclusions, as explained in Sec.~\ref{sec:real}. We hope that this line of research will eventually lead to the establishment of a connection or a systematic differentiation between dynamical and kinematic nonlocality.

In principle, the results presented here can be experimentally realized. In this regard, changes associated with modular variables in interferometers were already observed in different platforms~\cite{spence2012experimental, carvalho2012experimental, fluhmann2018sequential}. Also, an experiment has been proposed to measure the alteration in modular kinetic momentum in an AB loop~\cite{kaufherr2014test}. Furthermore, experimental assessment of realism within interferometers in the study of wave-particle duality has been recently conducted~\cite{dieguez2022experimental}. A similar approach, in an appropriate platform, can be used to evaluate the results of this article.

It is also noteworthy that other geometric and topological phases~\cite{berry1984quantal, anandan1990geometry, cohen2019geometric, paiva2022geometric}, including, e.g., the Aharonov-Casher effect~\cite{aharonov1984topological}, can also be analyzed using the same method as the one in this work. However, as we have already seen, this approach is particularly relevant in cases involving relative phases that are gauge dependent.

Moreover, to focus on fundamental aspects, we have considered here a relatively simple scenario with the AB effect. However, one may also extend these calculations to the case where different amplitudes travel on each arm of the interferometer. While this would generally modify the values computed here, the abrupt change would remain a characteristic of the realism of a charge in the AB scenario. Furthermore, in experimental setups, it might be necessary to take into account the contribution of other effects at play in addition to the AB effect. For instance, Ref.~\cite{noguchi2014aharonov} investigated the tunneling of a charge coupled to a vector potential. Then, in addition to the AB effect, the contribution of the potential associated with tunneling would be relevant to the analysis. However, the main results we have obtained and in particular the conclusion about the discontinuous change in realism will still hold. The difference is that extra contributions should be added to the computation of realism.

Finally, it would be interesting to consider how models for the quantized electromagnetic field, investigated, e.g., in Refs.~\cite{marletto2020aharonov, saldanha2021shielded}, modify our conclusions about the realism of the observables considered here. However, these models include a Fock space and the framework of realism assumes finite-dimensional systems. Nevertheless, special constructions by means of coarse graining can be made to study the realism of continuous-variable systems~\cite{freire2019quantifying, lustosa2020irrealism}. Then an investigation of these models with quantized electromagnetic fields requires, first, the development of an appropriate framework to analyze realism in Fock spaces, which is beyond the scope of the present work. \\

The code to generate the graphs presented here can be found in Ref.~\cite{paiva2022code}.

\acknowledgements{This research was supported by Grant No. FQXi-RFP-CPW-2006 from the Foundational Questions Institute and Fetzer Franklin Fund, a donor-advised fund of Silicon Valley Community Foundation. I.L.P. acknowledges financial support from the ERC Advanced Grant FLQuant. P.R.D. acknowledges support from the Foundation for Polish Science (IRAP project, ICTQT, Contract No. MAB/2018/5, co-financed by EU within Smart Growth Operational Programme). R.M.A. acknowledges support from CNPq/Brazil (Grant No. 309373/2020-4) and the National Institute for Science and Technology of Quantum Information (CNPq, INCT-IQ 465469/2014-0). E.C. was supported by the Israeli Innovation Authority under Projects No. 70002 and No. 73795, the Pazy Foundation, the Israeli Ministry of Science and Technology, and the Quantum Science and Technology Program of the Israeli Council of Higher Education.}

\appendix

\section{Computation of realism of $\hat{\sigma}_g(\theta)$}
\label{app:std-int}

In this appendix we give details of the calculation of the realism of $\hat{\sigma}_x$ and $\hat{\sigma}_g(\theta)$ associated with the state $\ket{\psi_0(\theta)}$ in Sec.~\ref{sec:std-int}. This result will also be useful for the analysis of the realism of other quantities considered in Sec.~\ref{sec:std-ab}.

To start, observe that $\hat{\sigma}_x$ belongs to the family of $\hat{\sigma}_g(\theta)$ indexed by functions $g$ with $g\equiv0$. Then, our calculations can be focused in the realism of $\hat{\sigma}_g(\theta)$.

Since $\mathfrak{R}_{\hat{\sigma}_g(\theta)}(\hat{\rho}_S) = 1 - S(\Phi_{\hat{\sigma}_g(\theta)}(\hat{\rho}_S))$, we need to compute $\Phi_{\hat{\sigma}_g(\theta)}(\hat{\rho}_S)$. For this, first note that the eigenvectors of $\hat{\sigma}_g(\theta)$ are
\begin{equation}
    \ket{x_g^0(\theta)\pm} = \frac{1}{\sqrt{2}} \left(\ket{0} \pm e^{ig(\theta)} \ket{1}\right).
\end{equation}
As a consequence,
\begin{equation}
    \begin{aligned}
        |\psi_0(\theta)\rangle = \frac{1}{2} &\left[\left(1+ e^{i[f(\theta) - g(\theta)]}\right) \ket{x_g^0(\theta)+} \right. \\
            &\left. + \left(1- e^{i[f(\theta) - g(\theta)]}\right) \ket{x_g^0(\theta)-} \right].
    \end{aligned}
\end{equation}
Therefore, $S(\Phi_{\hat{\sigma}_g(\theta)}(\hat{\rho}_S)) = h(\lambda)$ with $\lambda = \{1 + \cos[f(\theta) - g(\theta)]\}/2$ and $\mathfrak{R}_{\hat{\sigma}_g(\theta)}(\hat{\rho}_S) = 1 - h(\lambda)$ as stated in Section~\ref{sec:std-int}. Finally, we see that, in particular, $\mathfrak{R}_{\hat{\sigma}_x}(\hat{\rho}_S) = 1 - h(\lambda)$ with $\lambda = \{1 + \cos[f(\theta)]\}/2$.

\section{Computation of realism of $\hat{\Sigma}_x(\theta)$}
\label{app:q-ab-x}

Here we show the steps for the derivation of the realism of $\hat{\Sigma}_x(\theta)$, which is defined in Eq.~\eqref{eq:def-sigma-quant}. First, observe that, for every $m_\ell$, its eigenvectors can be written as
\begin{equation}
    \ket{X_f^{m_\ell}(\theta)\pm} = \ket{x_f^{m_\ell}(\theta)\pm} \otimes \ket{m_\ell},
\end{equation}
with
\begin{equation}
    \ket{x_f^{m_\ell}(\theta)\pm} \equiv \frac{1}{\sqrt{2}} \left(\ket{0} \pm e^{i \left[f(\theta) - q\int_{\tau(\theta)} \vec{A}_{m_\ell} \cdot d\vec{s}/\hbar\right]} \ket{1}\right).
\end{equation}
Then,
\begin{equation}
    \Phi_{\hat{\Sigma}_x(\theta)}(\hat{\rho}) = \sum_{m_\ell\in\Gamma_\ell} |c_{m_\ell}|^2 \hat{\rho}_S^{m_\ell}(\theta) \otimes \ket{m_\ell} \bra{m_\ell},
\end{equation}
where
\begin{equation}
    \begin{aligned}
        \hat{\rho}_S^{m_\ell}(\theta) \equiv \frac{1}{2} & \left\{\left[1 + \cos(q\Phi_{enc}^{m_\ell}/\hbar)\right] \ket{x_f^{m_\ell}(\theta)+} \bra{x_f^{m_\ell}(\theta)+} \right. \\
            &\left. + \left[1 - \cos(q\Phi_{enc}^{m_\ell}/\hbar)\right] \ket{x_f^{m_\ell}(\theta)-} \bra{x_f^{m_\ell}(\theta)-}\right\}.
    \end{aligned}
\end{equation}

Using the joint entropy theorem~\cite{nielsen2000quantum}, we conclude that $S(\Phi_{\hat{\Sigma}_x(\theta)}(\hat{\rho})) = h(\vec{C}) + \sum_{m_\ell\in\Gamma_\ell} |c_{m_\ell}|^2 h(\lambda_{m_\ell}(\theta))$, where $\vec{C} = (|c_{m_\ell}|^2)$ and $\lambda_{m_\ell}(\theta) = [1 + \cos(q\Phi_{enc}^{m_\ell}/\hbar)]/2$.

Finally, we write $\mathfrak{R}_{\hat{\Sigma}_x(\theta)}(\hat{\rho}) = 1 - h(\vec{C}) - \sum_{m_\ell\in\Gamma_\ell} |c_{m_\ell}|^2 h(\lambda_{m_\ell}(\theta))$. Observe that, if $\theta < \pi/2$, the last term vanishes since $\lambda_{m_\ell}(\theta) = 1$ for every $m_\ell$. This changes if $\theta > \pi/2$ since, in this case, $\Phi_{enc}^{m_\ell} = \Phi_B^{m_\ell}$.

\section{Computation of the realism of $\hat{\sigma}_z$}
\label{app:sigma-z}

In this appendix we calculate the realism of $\hat{\sigma}_z$ for the joint state $\hat{\rho}$ of the charge and the cylinder, as discussed in Sec.~\ref{sec:q-ab}. For this we need to use that
\begin{equation}
    \Phi_{\hat{\sigma}_z}(\hat{\rho}) = \frac{1}{2} \left(\ket{0} \bra{0} \otimes \ket{\xi'} \bra{\xi'} + \ket{1} \bra{1} \otimes \ket{\xi''} \bra{\xi''}\right),
\end{equation}
where
\begin{equation}
    \ket{\xi'} = \sum_{m_\ell \in \Gamma_\ell} c_{m_\ell} e^{ig_{m_\ell}(\theta)} \ket{m_\ell}.
\end{equation}
and
\begin{equation}
    \ket{\xi''} = \sum_{m_\ell \in \Gamma_\ell} c_{m_\ell} e^{i[g_{m_\ell}(\theta) + Kq m_\ell \theta/ \pi]} \ket{m_\ell}.
\end{equation}
Then, using the joint entropy theorem~\cite{nielsen2000quantum}, we conclude that $S(\Phi_{\hat{\sigma}_z}(\hat{\rho})) = \log 2$ and hence $\mathfrak{R}_{\hat{\sigma}_z}(\hat{\rho}) = \log(4\ell+2) - S(\Phi_{\hat{\sigma}_z}(\hat{\rho})) = \log(2\ell+1)$ for every $\theta$. With base-$(4\ell + 2)$ logarithms, this value is $\log_{4\ell+2}(2\ell+1) = 1 - \log_{4\ell+2}2$. However, if base-$2$ logarithms are used, like in Eq.~\eqref{eq:def-real}, we conclude that $\mathfrak{R}_{\hat{\sigma}_z}(\hat{\rho}) = 0$.

\bibliography{bibliography}

\end{document}